\documentclass[aps,pre,citeautoscript,twocolumn,superscriptaddress,nofootinbib,nobibnotes,10pt]{revtex4-2} 
\usepackage[utf8]{inputenc}
\usepackage{amsmath,amssymb}
\usepackage{mathrsfs} 
\usepackage{mathtools}
\usepackage{siunitx} 
\usepackage{graphicx,color}
\usepackage{varwidth,xcolor}
\usepackage{array} 
\usepackage{placeins}
\usepackage{ifthen}
\usepackage{hyperref}\usepackage[capitalise]{cleveref}
\usepackage{natbib} 
\usepackage[normalem]{ulem} 

\DeclareGraphicsExtensions{.pdf,.PDF,.png,.PNG}

\definecolor{myblue}{RGB}{20,80,150} 	
\definecolor{myred}{RGB}{160,30,30} 	
\definecolor{mygreen}{RGB}{50,120,50} 	

\hypersetup{
	colorlinks=true,
	linkcolor=myblue, 	
	citecolor=myred, 	
	urlcolor=mygreen 	
}

\usepackage{braket}
\usepackage{mathtools}
\DeclareMathAlphabet\mathbfcal{OMS}{cmsy}{b}{n}

\begin{document}
	
\title{Thermo-optical spiking and mixed-mode oscillations in injected Kerr microcavities}

\author{Elias R. Koch}
\email[Corresponding author: ]{elias.koch@uni-muenster.de}
\affiliation{Institute for Theoretical Physics, University of M\"unster, Wilhelm-Klemm-Str. 9, 48149 M\"unster, Germany
}%

\author{Julien Javaloyes}
\affiliation{Departament de F\'{\i}sica, Universitat de les Illes Balears \& IAC-3, Cra.\,\,de
	Valldemossa, km 7.5, E-07122 Palma de Mallorca, Spain
}%
\author{Svetlana V. Gurevich}
\affiliation{Institute for Theoretical Physics, University of M\"unster, Wilhelm-Klemm-Str. 9, 48149 M\"unster, Germany
}%
\affiliation{Center for Nonlinear Science (CeNoS), University of M\"unster, Corrensstraße 2, 48149 M\"unster, Germany
}%

\date{\today}

\begin{abstract}
We investigate the nonlinear dynamics of vertically emitting Kerr microcavities under detuned optical injection, considering the impact of slow thermal effects. Our model integrates thermal detuning caused by refractive index shifts due to heating. Through numerical and analytical approaches, we uncover a rich spectrum of dynamical behaviors, including excitable thermo-optical pulses, mixed-mode oscillations, and chaotic spiking, governed by a higher-dimensional canard scenario. Introducing a long external feedback loop with time delays comparable to the microcavity photon lifetime but shorter than thermal relaxation timescales, reveals how delay affects excitability and stabilizes temporal localized states. Our findings extend the understanding of excitable systems, demonstrating how thermal and feedback mechanisms interplay to shape nonlinear optical dynamics. Further, our approach paves the way for the study of cavity stabilization and cavity cooling using an additional control beam.
\end{abstract}

\maketitle
\section{Introduction}

The phenomenon of excitability is a key concept across various research fields that has attracted considerable attention over the last decades~\cite{Winfree1980,Meron19921,Izhikevich,MS-PhysRep-06,Mikhailov2012}. A list of examples includes biological systems such as neurons in the human brain~\cite{I_IJBC_00}, chemical reactions like the Belousov-Zhabotinsky reaction~\cite{KAK_Nature_89}, neuro-inspired computing~\cite{PSL_AOP_16, BPV_PhysRevRes_24}, cardiac arrhythmias~\cite{DPS_Nature_92, QHG_PhysRep_14}, fluid systems such as faraday waves ~\cite{HKV_PhysD_2005,HPK_Chaos_2008}, lasers~\cite{PVA_EPL_97} or extreme weather events \cite{AKLF_PRE_13}, to name just a few. There, excitations can be found in a purely temporal domain, however, spatially propagating excitations such as spiral waves, observed e.g., in the context of cardiac arrhythmias~\cite{DPS_Nature_92}, liquid crystals~\cite{CFG_Chaos_94}, or two-dimensional vertical-cavity surface-emitting lasers (VCSELs)~\cite{PHBH_PRL_05} can be found. 

\begin{figure}[t!]
	\centering
	\includegraphics[width=1\columnwidth]{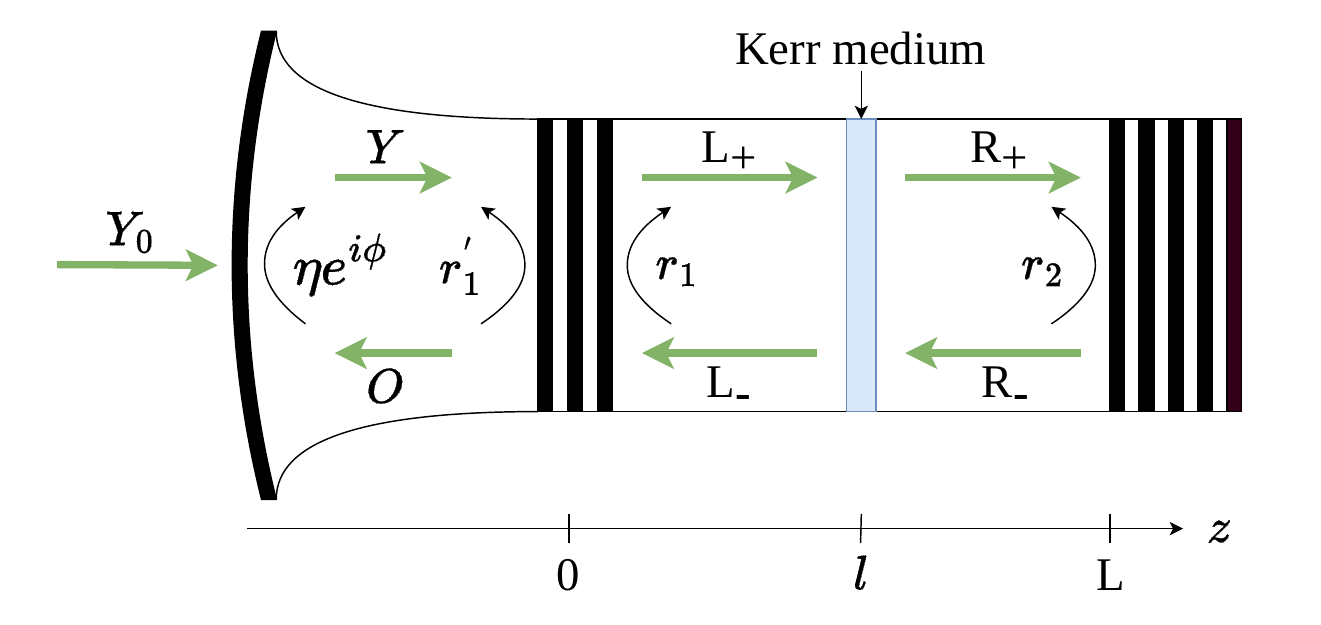}
	\caption{Schematic of a micro-cavity containing a Kerr medium at $z=l$ closed by distributed Bragg mirrors at $z=0$ and $z=L$. It is subjected to optical feedback from an external cavity closed by a mirror with reflectivity $\eta$ and phase $\phi$. The compound cavity system is driven by a continuous wave injection beam with amplitude $Y_0$. The reflection coefficients are indicated by $r_{1,2}$ for the left and right mirrors, respectively, whereas the prime mark reflection and transmission processes originating from outside of the  micro-cavity. The intracavity field is split into left $L_\pm$ and right $R_\pm$ counter-propagating waves on each side of the the Kerr slice.}\label{fig:1}
\end{figure}

In this paper, we investigate the influence of thermal effects on the dynamics of a vertically emitting Kerr-micro-cavity operated in the Gires–Tournois regime (KGTI)~\cite{GT-CRA-64,SPV-OL-19,SJG-OL-22,SGJ-PRL-22,KSGJ_OL_22,SKJGW_Chaos_23} that is subjected to strong time-delayed optical feedback and detuned optical injection, see  Fig.~\ref{fig:1}. To this aim, we derive a delay algebraic equations (DAEs) model including the effects of thermal expansion and temperature dependent refractive index. 
In particular, we report and analyze the occurrence of \emph{excitable thermo-optical pulses} (TOPs) in a higher-dimensional canard scenario, featuring  relaxation- and mixed-mode oscillations (MMOs)~\cite{SFM_PRE_05,MM_PRE_13, ZB_NonlinearDyn_15}.
We find that temporal localized states with  a period of approximately one round-trip in the external cavity, previously observed in the KGTI system without thermal effects~\cite{SPV-OL-19,SJG-OL-22} still manifest themselves as stable periodic solutions, while also TOPs and relaxation oscillations can be observed, featuring oscillatory details that evolve on the timescale of the round-trip in the external cavity.

In the framework of optical systems, excitability was studied theoretically and observed experimentally in e.g., lasers with optical feedback~\cite{GGG_PRE_97, EM_PRE_99,WBRH_PRL_01,PHBH_PRL_05, DVDMB_JPhysPhotonics_21}, injected lasers \cite{CDT_PRE_98,WBRH_PRL_01,MB_PRL_05,WLS_PhysRep_05}, lasers with saturable absorber~\cite{PVA_EPL_97,BKY_OL_11,SBB_PRL_14}, Fabry-Perot cavities~\cite{MDM_PRE_06,JB_IEEE_07}, semiconductor ring laser systems ~\cite{BMG_PhysLettA_10} and all-fiber lasers \cite{OGB_JOSAB_21}. In particular, excitability manifests itself in studies of the so-called low-frequency fluctuations observed in the laser output. These fluctuations appear as abrupt decreases in intensity almost to zero, followed by a slow recovery \cite{GGG_PRE_97,EM_PRE_99,BPG_PRE_03}. Additionally, a canard scenario corresponding to the excitability transition in the van der Pol-FitzHugh-Nagumo model has been observed in the thermo-optical dynamics of semiconductor optical amplifiers (SOAs) and VCSELs~\cite{BPG_PRE_03, MCSB_PRL_04, SK_JOSAB_11}. Here, excitability occurs due to the major thermal effects in optical microcavities: thermal expansion~\cite{MDM_PRE_06} and temperature dependent refractive index~\cite{DSBP_NatPhotonics_20}. Both effects mentioned enter the dynamics in a similar way by changing the effective cavity length, affecting the microcavity resonance and ultimately the detuning. As these effects are significantly slower than the other intrinsic timescales of the optical systems, they introduce a drastic scale separation~\cite{BPV_PhysRevRes_24} responsible for the observation of excitability. As such, this work on thermal effects provides an important first step towards the investigation of thermal cavity stabilization and cavity cooling, which can be approached using an additional control beam and a secondary resonance as discussed in more detail in ~\cite{DSBP_NatPhotonics_20}.
In general, a dynamical system is considered to be excitable if its unique stable steady state, in this context also referred to as the quiescent state~\cite{I_IJBC_00}, may respond to an external perturbation in two different ways: If the perturbation amplitude surpasses a certain small threshold, the system responds with a macroscopic excursion through the phase space. Alternatively, it relaxes exponentially to the quiescent state in response to perturbations below the threshold. The large amplitude response, whose shape depends only weakly on the details of the initial perturbation, is followed by a refractory time, during which the system ignores further perturbations. Excitable orbits typically occur in systems with at least two well-separated timescales.
The presence of excitability is usually a sign of proximity to a region of self-sustained relaxation oscillations whose shape and amplitude are related to that of the excitable orbits. The respective bifurcation scenarios, that connect the quiescent state to the relaxation oscillations, are particularly interesting, since they have to contain a sharp transition to the large-amplitude orbits. The most common scenarios are a saddle–node bifurcation on an invariant circle (SNIPER, or Andronov bifurcation) or an Andronov-Hopf (AH) bifurcation \cite{I_IJBC_00}. In the AH scenario, one might expect a subcritical bifurcation and, indeed, this is a scenario in the Hodgkin-Huxley model for neural excitability~\cite{I_IJBC_00} or resonant tunneling diodes~\cite{OrtegaPiwonka2021}. However, excitability can also occur in a supercritical AH scenario, well known from the van der Pol–FitzHugh-Nagumo model~\cite{FH_BiophysJ_61, N_IEEE_62}, which is a simplified model for neuron dynamics. Here, a narrow regime of subthreshold oscillations emerges at the primary bifurcation \cite{MMB_PRL_07}, followed by an abrupt though continuous increase of amplitude and period. This phenomenon, known as a canard explosion, leads to the relaxation oscillation regime. Note that subthreshold oscillations can still exhibit excitations if perturbed above the respective threshold~\cite{MNV_PRL_01}. This comparably simple scenario drastically changes if a third variable, such as inertia, is added to the system~\cite{MMB_PRL_07}  leading to new effects that disrupt the clear separation between subthreshold and relaxation oscillations.

\section{Model System}
The schematic setup is depicted in Fig.~\ref{fig:1}. It is composed of a monomode micro-cavity of a few micrometers between two Bragg mirrors located at $z=0$ and $z=L$. The corresponding reflection coefficients are $r_{1,2}$ for the top and the bottom mirror, respectively, whereas primes denote the case on the outside of the micro-cavity. The intracavity field is split into left $(-)$ and right $(+)$ moving waves to the left ($L$) and right ($R$) of thin slice of Kerr material that is situated at the field's anti-node at $z=l$. The micro-cavity is coupled with a long external cavity, (typically) a few centimeters in length, with a round-trip time $\tau$. The external cavity is closed by a feedback mirror with reflectivity $\eta$ and phase $\phi$. The system is driven by continuous wave (CW) injection with amplitude $Y_0$ and frequency $\omega_0$. The light coupling efficiency in the cavity is
given by the factor $h=h(r_1,r_2)$, where the choice of a perfectly reflecting bottom mirror ($r_2=1$) results in $h=2$, which corresponds to the Gires-Tournois interferometer regime \cite{GT-CRA-64}. A Gires-Tournois interferometer is essentially an asymmetrical Fabry-Perot resonator, composed of one highly reflective mirror and one semi-transparent mirror. The latter are widely used as optical pulse-shaping elements, inducing a controllable amount and sign of group delay dispersion, which is the dominating effect outside resonance.

The optical properties of the cavity change with temperature. Beyond the direct thermal expansion of the crystal, the thermo-optical effect represents the change in refractive index  with temperature \cite{DCI-APL-00}. Both effects lead to a change of the cavity resonant frequencies $\omega_c$. The direction and the strength of this effect depends on the material itself. For a microcavity injected with a field of amplitude $Y_0$ and frequency $\omega_0$, the microcavity resonance is incorporated into the model through the detuning $\delta=\omega_c-\omega_0$. As we include thermal effects, $\omega_c$ becomes a time-dependent variable through the evolution of the temperature, driven by optical absorption acting as a heat source. Following the lines of~\cite{JB_IEEE_07,SPV-OL-19} we can derive the model equations (see App.~\ref{appendixA} for details) as
\begin{align}
\dot{E}= & \left[-1+i\left(s|E|^{2}-\delta\right)\right]E+hY,\label{eq:Eq1}\\
\dot{\delta}= & \gamma\left(\delta_0-\delta-\mu_{\alpha}\left|E\right|^{2}\right)\label{eq:Eq2},\\
Y= & \eta e^{i\varphi}\left[E(t-\tau)-Y(t-\tau)\right]+\sqrt{1-\eta^{2}}Y_{0}.\label{eq:Eq3}
\end{align}
Here, $E$ and $Y$ denote the slowly varying envelopes of the electric fields in the micro-cavity and the external cavity, respectively. The cavity-enhanced nonlinear coefficient $s$ is in general complex and its real and imaginary parts model the effects of the self-phase modulation and two-photon absorption, respectively. Further, $\delta_0$ denotes the intrinsic detuning with respect to the microcavity resonance in the absence of the heat sources. 
The thermal inertia of the cavity is denoted by $\gamma$ while $\mu_\alpha$ represents the thermal nonlinearity of the cavity, with $\mu_\alpha$ in $\mathrm{m}^2/V^2$. The latter depends on the thermal impedance, the absorption, as well as the thermo-optic coefficients. As in~\cite{JB_IEEE_07}, the net amount of absorbed light is obtained by performing an energy balance between the norm of the Poynting vectors associated with the incoming and emitted fields, see App.~\ref{appendixA} for details.
The field cavity enhancement can be conveniently scaled out using Stockes relations allowing the intracavity and external cavity fields, $E$ and $Y$, to be of the same order of magnitude. This leads to the simple
input-output relation $O = E - Y$. The coupling between the intra-cavity and the external
cavity fields is given in Eq.~\eqref{eq:Eq3} by a DAE where the effects of the external mirror and signal extraction (e.g., a beam-splitter or transmission through the mirror itself) are combined in the attenuation factor $\eta$. The total phase upon re-injection is $\varphi = \omega_0 \tau + \phi$ consisting of the accumulated phase per round-trip due to propagation and the feedback mirror phase $\phi$. Finally, throughout our simulations, we apply Gaussian white noise with an amplitude of $\sigma$ to the fast $E$-field to observe excitability. 
\section{Thermo-optical excitations for short external cavity}
\subsection{Excitability through manipulating the nullclines}
We note that the introduced thermal evolution given by Eq.~\eqref{eq:Eq2} is typically much slower than the round-trip time $\tau$ in the external cavity. Hence, to isolate the impact of thermal effects on the KGTI system, we first consider
the limit $\tau\rightarrow 0$. This limit corresponds to a simple injected Kerr micro-cavity, reducing the DAE system~\eqref{eq:Eq1}-\eqref{eq:Eq3} to a system of ordinary differential equations (ODEs)
\begin{align}
 \dot{E}= & \left[i\left(s|E|^{2}-\delta\right)+\frac{h\eta e^{i\varphi}}{1+\eta e^{i\varphi}}-1 \right]E+\frac{h\sqrt{1-\eta^2}}{1+\eta e^{i\varphi}}Y_0,\label{eq:Eq4}\\
\dot{\delta}= & \gamma\left(\delta_0-\delta-\mu_{\alpha}\left|E\right|^{2}\right)\label{eq:Eq5}.
\end{align}

  \begin{figure}[t!]
  	\centering
	\includegraphics[width=1\columnwidth]{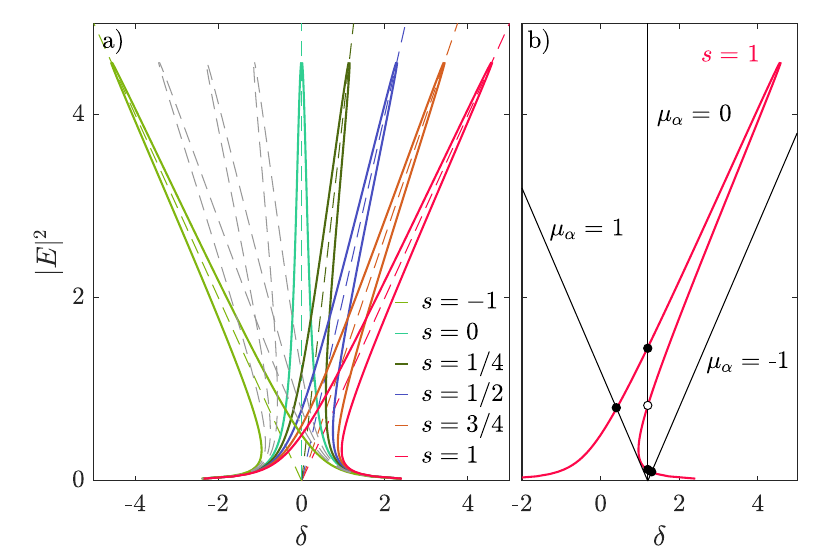}
 	\caption{a) A $E$-nullcline of Eqs.~\eqref{eq:Eq4},\eqref{eq:Eq5} for different values of $s$. The nullcline is bent away from the vertical position for $s=0$, enabling a bistable CW response in for the system. b)  $(E,\,\delta)$-nullcines for $s=1$ for different values of $\mu_\alpha$ (black). The  intercepts of both curves corresponds to stable (filled circle) and unstable (open circle) fixed-points of the CW solution. Parameters are: $(\delta_0,h,\eta,\varphi,Y_0,\gamma)=(1.2,2,0.7,0,0.45,0.01)$.}\label{fig:2}
\end{figure}
The zeros of the right-hand sides of Eqs.~\eqref{eq:Eq4},\eqref{eq:Eq5} define the so-called nullcline curves of the system and Fig.~\ref{fig:2}~(a) shows the $E$-nullcline for the intensity field $|E|^2$ of Eq.~\eqref{eq:Eq4} as a function of the thermal detuning $\delta$ for different values of $s$, that we consider here to be real for simplicity.
One can see that the nullcline without the Kerr effect (i.e. $s=0$) does not possess any folds, which excludes the possibility of bistability, whereas the bending of the nullcline curve away from the symmetric case $s=0$ equals the strength of the Kerr effect $s$.

\begin{figure}[t!]
	\centering
	\includegraphics[width=1\columnwidth]{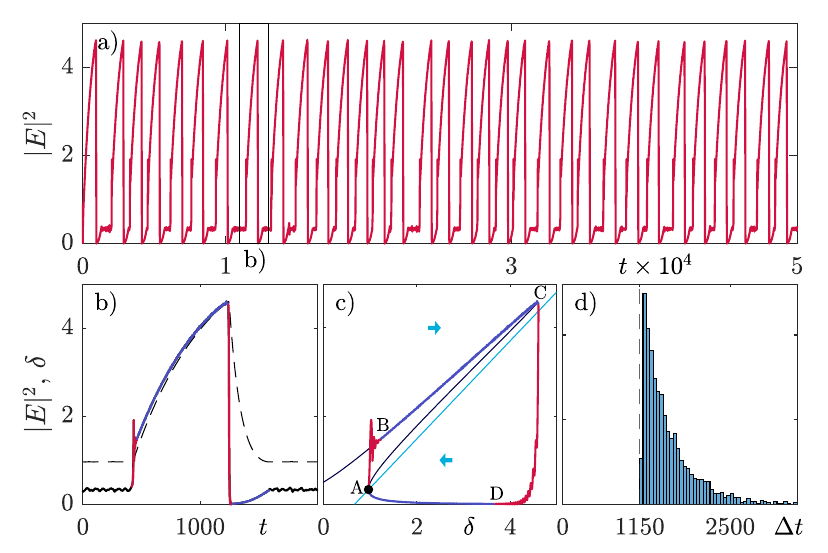}
	\caption{Excitable thermo-optical pulses on top of the stable continuous wave state. a) A typical numerical time trace of Eqs.~\eqref{eq:Eq4}-\eqref{eq:Eq5} with several excitations initiated by noise. b) Single excitation  with fast (red) and slow (blue) stages for the $E$ field, caused by multiple timescale dynamics. The $\delta$ field is depicted in dashed black. c) The same pulse in the $(\delta,|E|^2)$ projection of the phase space together with the $E$-nullcline (black) and the $\delta$-nullcline (light blue). d) The corresponding interspike interval histogram. Parameters are $(\delta_0,h,\eta,\varphi,Y_0,\mu_\alpha,\gamma,s)=(0.665,2,0.7,0,0.45,-0.9,0.01,1)$.
}
	\label{fig:3}
\end{figure}

Note that the full phase space $(\delta,\,\theta,\,|E|^2)$ of the system~\eqref{eq:Eq4}-\eqref{eq:Eq5} is three-dimensional,  where $\theta$ is the phase of the $E$-field. Hence, since~\eqref{eq:Eq2} is phase-invariant, the corresponding $\delta$-nullcline of Eq.~\eqref{eq:Eq5} is actually a plane. Figure~\ref{fig:2}~(b) presents the $E$-nullcline for $s=1$ together with the $\delta$-nullclines for three different values of $\mu_\alpha$ which directly determines the slope of the plane. One notices that by for example changing $\mu_\alpha$ and keeping $\delta_0$ fixed, the number, position, and stability of the fixed points resulting from the intercepts of the nullclines can be tuned.
We start by considering a parameter set corresponding to a single stable fixed point (cf. Fig. \ref{fig:2}~(b) for $\mu_\alpha=-1$) and fix the thermal scale $\gamma=0.01$, thus allowing for a time-scale separation between the field and the thermally controlled detuning.
We performed numerical simulations of Eqs.~\eqref{eq:Eq4}-\eqref{eq:Eq5} in the presence of noise. In this regime, the system exhibits TOPs on top of the stable CW state, as presented in Fig.~\ref{fig:3}. Here, panel (a) shows the field intensity of the resulting time trace with pronounced excitations initiated by noise. Figure~\ref{fig:3}~(b) contains a zoom into a single excitation pulse. Here, the dashed black line corresponds to the $\delta$-values and for the $E$-field profile one recognises the so-called fast (red) and slow (blue) stages, caused by multiple time-scale dynamics.

The origin of the slow and fast stages is illustrated in panel~(c), where the pulse profile is shown in the $(\delta,|E|^2)$ projection of phase space, along with the $E$-nullcline (black) and the $\delta$-nullcline (light blue). The stable fixed point marked as $A$ is at the intercept of both curves. A sufficiently strong perturbation can push the trajectory beyond the separatrix between the basin of attraction of the fixed point $A$ and the upper branch of the nullcline. Above the $\delta$-nullcline, there is only dynamics towards positive $\delta$ direction (blue arrow), however, since $\delta$ is a slowly evolving variable, the trajectory is rapidly pulled towards the upper branch of the $E$-nullcline in a fast stage $AB$, preventing a return to the fixed point. As the trajectory approaches the $E$-nullcline, the dynamics of the field $E$ is enslaved by the slow variable $\delta$, initiating the slow stage $BC$, where $\delta\approx|E|^2$ holds. At the upper fold of the $E$-nullcline, point $C$, the trajectory can no longer follow the nullcline because the phase space region above the $\delta$-nullcline is constrained to flow in the positive $\delta$ direction. Hence, the trajectory departs from the nullcline, initiating the fast stage $CD$, characterized by oscillatory dynamics to be discussed later. Finally, near point $D$, the trajectory reaches the lower branch of the $E$-nullcline, starting another slow stage $DA$ that returns to the fixed point $A$. Panel (d) shows the corresponding interspike-interval (ISI) histogram ~\cite{LNK_PRE_98}. There, one can see the characteristic cutoff at $\Delta t\approx 1150$ due to the refractory time and an exponential decay in the likelihood of longer interspike intervals. For this and all subsequent ISI histograms, the interspike intervals of a time trace featuring approximately $6000$ excitations were evaluated by calculating all peak-to-peak distances and sorting them into discrete, equidistant bins. It is normalized in a way that all bars sum up to one. \\
\begin{figure}[t!]
	\centering
	\includegraphics[width=1\columnwidth]{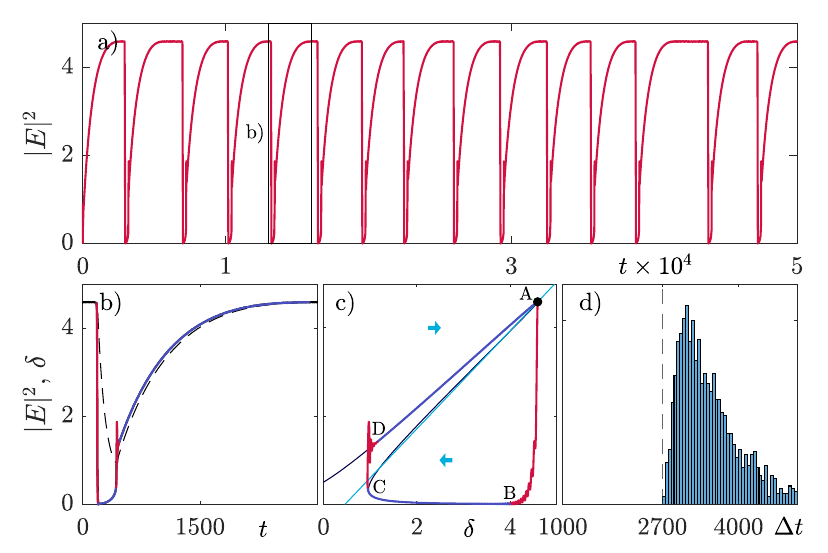}
	\caption{Dark excitable thermo-optical pulses on top of the upper stable continuous wave state obtained by numerical integration of Eqs.~\eqref{eq:Eq4}-\eqref{eq:Eq5} for $\delta=0.46$. a) A typical numerical time trace with several excitations initiated by noise. b) Single excitation with fast (red) and slow (blue) stages, caused by multiple timescale dynamics. The $\delta$ field is depicted in dashed black. c) The same pulse in the $(\delta,|E|^2)$ projection of the phase space together with the $E$-nullcline (black) and the $\delta$-nullcline (light blue). d) The corresponding interspike interval histogram. Other parameters as in Fig.~\ref{fig:3}.
}
	\label{fig:4}
\end{figure} 
A slight reduction in $\delta_0$ can shift the fixed point $A$ to the upper fold, see Fig.~\ref{fig:4}~(c) where the system is monostable once again. In the presence of noise, we now observe dark TOP excitations, or dropouts, as depicted in Fig.~\ref{fig:4}~(a,b), occuring on the upper CW value. These excitations have a longer refractory time, resulting in a larger cutoff in the ISI histogram, as shown in Fig.~\ref{fig:4}~(d). This is due to the $\delta$-nullcline being closer to the slow stage $DA$. Additionally, the ISI histogram appears to exhibit a smoothed cutoff, likely due to the noise amplitude being below the coherence resonance. According to the concept of coherence resonance, there exists an optimal noise amplitude leading to an almost periodic train of excitations, with adjacent excitations occurring nearly immediately after recovery. In this case, the ISI histogram displays a maximum peak near the recovery time, followed by an exponential decay. Note that the ISI histogram of a periodic solution contains only a single peak at the corresponding period.

Interestingly, the refractory time for dark excitations is significantly longer than that for bright excitations. This is primarily due to the prolonged slow stage along the upper branch of the $E$-nullcline, caused by the reduced distance to the left-shifted $\delta$-nullcline in the dark TOP case, reducing the velocity of this slow stage even further. Consequently, the other slow stage is shorter in the dark case. However, since this stage is generally relatively short for dark and bright excitable TOPs alike, the total excitation period is still increased.
\subsection{Enslaved and unbound phase dynamics}
\begin{figure}[t]
	\centering
	\includegraphics[width=1\columnwidth]{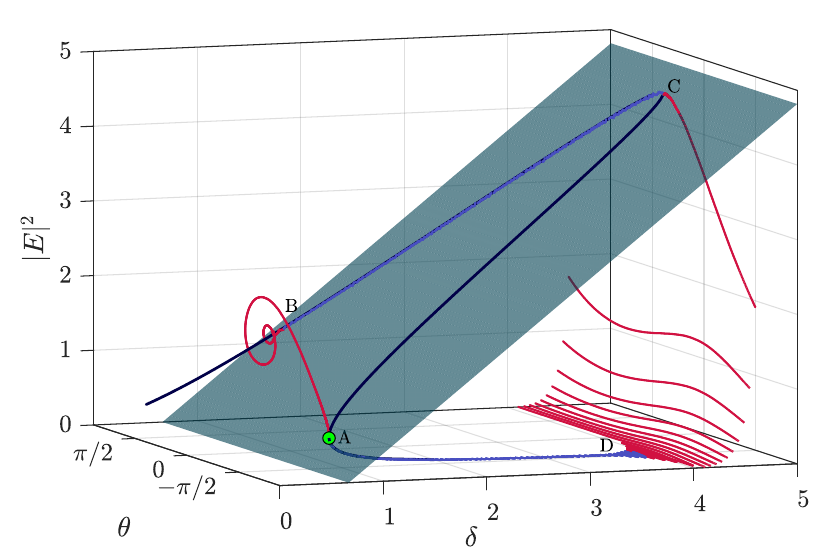}
	\caption{The trajectory of the bright excitable TOP (cf. Fig.~\ref{fig:3}) (red for fast ($AB$, $CD$) and blue for slow ($BC$, $DA$) stages, respectively) in the three-dimensional phase space $(\delta,\,\theta,\,|E|^2,)$. The $E$-nullcline is given by a dark blue line , as the real and imaginary parts of $E$ have a fixed relation on the nullcline. The $\delta$-nullcline does not depend on the phase $\theta$, hence it is a plane in the phase-space. The green circle at $A$ marks the single stable fixed point located at the intercept of the nullclines. Parameters as in Fig.~\ref{fig:3}.
}
	\label{fig:5}
\end{figure}

Figure~\ref{fig:5} illustrates the full three-dimensional phase space $(\delta,\, \theta,\,|E|^2)$ of Eq.~\eqref{eq:Eq4} (dark blue) and Eq.~\eqref{eq:Eq5} (shaded plane). The green circle $A$ indicates the single stable fixed point located at the intercept of the nullclines.
Additionally, the trajectory of a bright TOP, as depicted in Fig.~\ref{fig:3}~(b), is represented here with red and blue segments denoting fast and slow stages of the dynamics, respectively.

The full phase dynamics along the excitation trajectory reveal rotational motion in the $(|E|^2, \theta)$ plane. This phase rotation is primarily influenced by the term $i(-\delta+|E|^2)$ from Eq.~\eqref{eq:Eq4} for $s=1$. During the slow stage $BC$, $\delta\approx |E|^2$ holds, as visible in Fig.~\ref{fig:3}~(b), causing the phase rotation to vanish, i.e. the phase is enslaved. Upon reaching point $C$, where $\delta_C\approx |E_C|^2$, the intensity rapidly decreases during the adjacent fast stage.
Owing to the time-scales separation governed by $\gamma$, the detuning $\delta$ remains approximately constant at a value $\delta_C$ during the relaxation of $E$. Consequently, the rotational velocity increases towards $i(-\delta_C)$, causing the rise in oscillation speed in negative $\theta$-direction observed in the fast stage $CD$ in Fig.~\ref{fig:5}.

The fast stage ends as the trajectory approaches the lower branch of the $E$-nullcline, initiating a second slow recovery stage $DA$ that returns to the fixed point. Since the maximum rotation is proportional to $\delta_0$, the oscillations occurring along the fast stage can be tuned by selecting a parameter regime with smaller $\delta_0$. As $\delta_0$ represents the detuning between the micro-cavity resonance without thermal heating $\omega_c$ and the injection frequency $\omega_0$, it can be modified by tuning the external injection frequency.
\subsection{Accessing different dynamical regimes}
As already discussed above, the form of the $E$-nullcline is influenced by the parameters $\eta,\, \varphi,\,Y_0$ as well as the strength of the Kerr effect $s$. However, the way these parameters enter the nullcline is not trivial. This is different for the $\delta$-nullcline following the simple equation $\delta=\delta_0-\mu_\alpha|E|^2.$
\begin{figure}[t!]
	\centering
	\includegraphics[width=1\columnwidth]{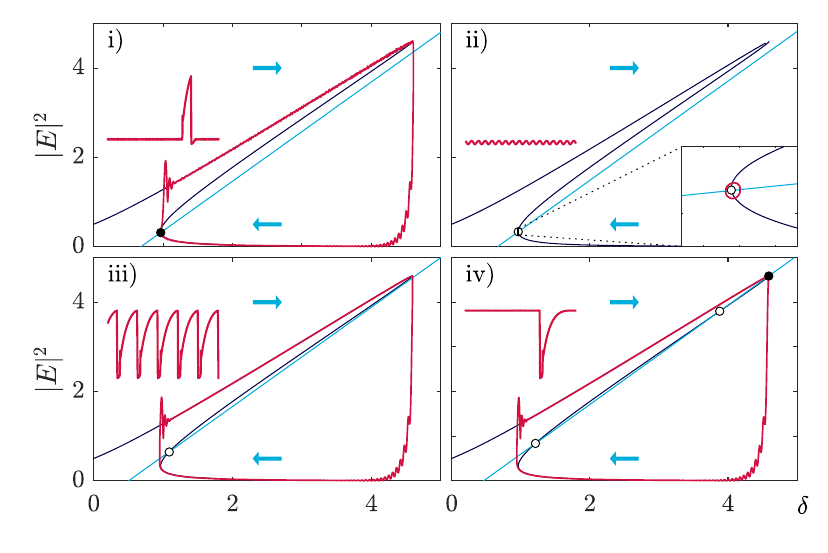}
	\caption{Different dynamical regimes obtained numerically from  Eqs.~(\ref{eq:Eq4},\,\ref{eq:Eq5}) (red line) in the $(\delta,\,|E|^2)$ phase space for four different values of $\delta_0$: (i) bright TOP; (ii) subthreshold oscillations; (iii) relaxation oscillations; (iv) dark TOP. The $E$- and $\delta$ nullclines are depicted in dark and light blue, respectively. Black filled (open) circle mark stable (unstable) fixed point. $\sigma$ indicates the amplitude of the Gaussian white noise. Other parameters as in Fig.~\ref{fig:3}.
}
\label{fig:6}
\end{figure} 
One can observe that $\delta_0$ is the $\delta$-intercept while $\mu_\alpha$ represents the slope. By tuning the injection frequency, i.e. $\delta_0$, we are able to shift the $\delta$-nullcline and change the behavior of the system in question. An overview over four different dynamical regimes that can be accessed by tuning a single parameter is shown in Fig.~\ref{fig:6}. Here, each panel depicts the $(\delta,\,|E|^2)$ projection of the phase space together with the $E$-nullcline (dark blue) and the $\delta$-nullcline (light blue). The blue arrows indicate the flow direction induced by the $\delta$-nullcline. Note that it can be drawn into the two-dimensional projection, since the corresponding nullcline is invariant with respect to the phase. For the $E$-nullcline this is different, since the flow strongly depends on the explicit value of the phase, for example leading to oscillation around the one-dimensional $E$-nullcline. However, the flow direction induced by the $\delta$-nullcline together with the position and number of stable (black filled circle) and unstable (black open circle) fixed points already allows to predict the systems behavior. In particular, Fig.~\ref{fig:6}~(i) shows the scenario of a single stable fixed point close to the lower fold of the $E$-nullcline. As discussed in Fig.~\ref{fig:3}, time simulation devoid any noise or other perturbations will end and stay on this fixed point. However, bright excitable TOPs (red curve) are possible by adding noise to the $E$-field of Eq.~\eqref{eq:Eq4}. If $\delta_0$ is slightly decreased, the fixed point is losing its stability and subthreshold oscillations can arise in a canard scenario close to the primary supercritical AH bifurcation, see Fig. \ref{fig:6}~(ii). Further, it is possible to excite the system on top of the subthreshold oscillation, if it is perturbed above the threshold and if $\delta_0$ is further reduced, a sharp transition of period and amplitude is observed, leading to a regime of self-sustained relaxation oscillations, which are shown in panel (iii). At some point, for even smaller values of $\delta_0$, a pair of stable and unstable fixed-points appears around the top fold of the $E$-nullcline as shown in panel (iv). Here, dark excitable TOP set in if noise is applied, cf. Fig.~\ref{fig:4}. The further the stable fixed point departs from the fold, the stronger perturbations are necessary to produce these excitations. For an even smaller value of $\delta_0$, the stable fixed point is no longer in the bistable regime and vertical perturbations through noise applied to Eq.~\eqref{eq:Eq4} cannot produce any hysteresis behavior anymore.
\begin{figure}[t!]
	\centering
	\includegraphics[width=1\columnwidth]{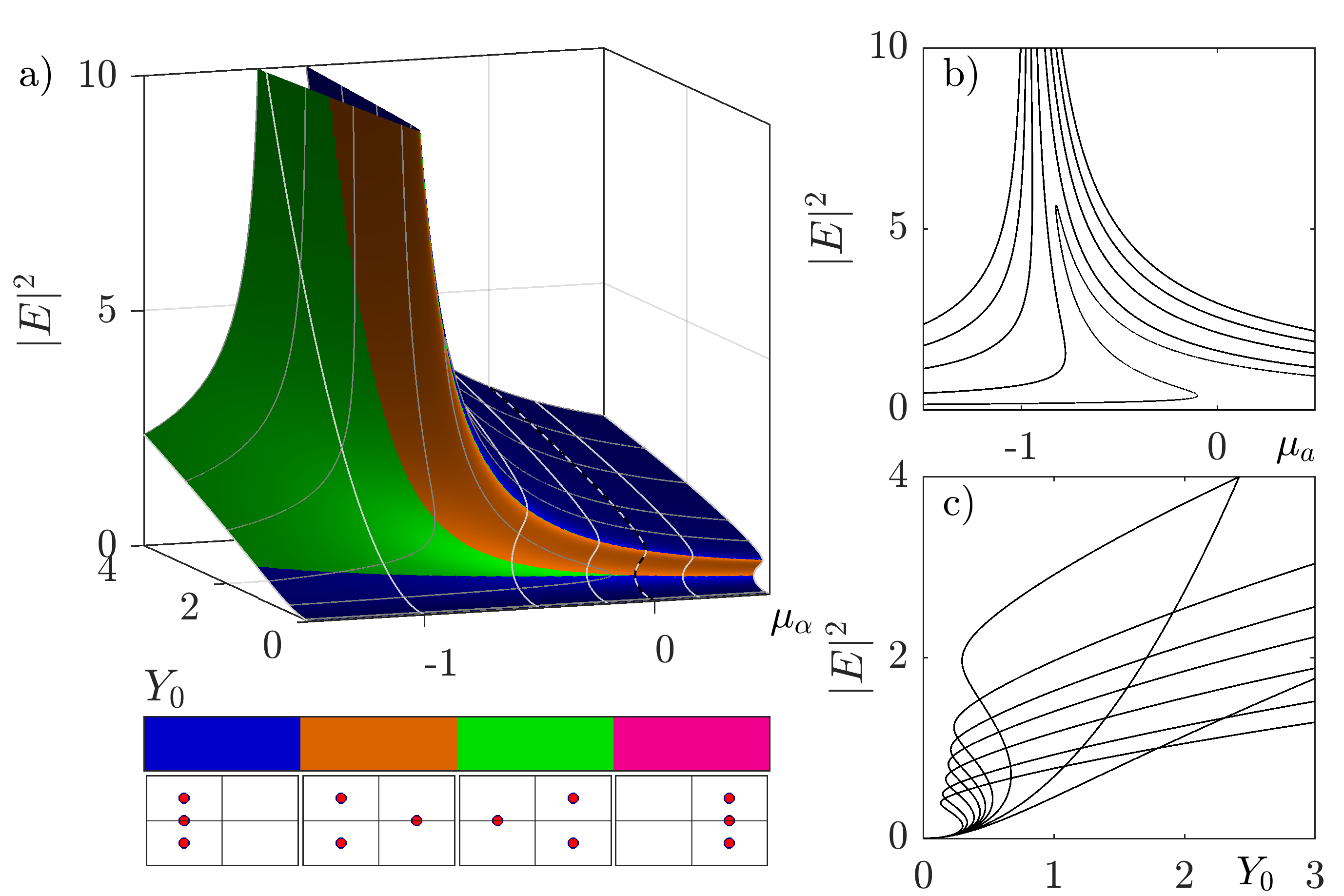}
	\caption{(a) CW solution of Eqs.~(\ref{eq:Eq4},\,\ref{eq:Eq5}) in the  the $(\mu_\alpha,\, Y_0)$ plane. The colors encode the stability information as indicated in the legend at the bottom: blue, orange,green and margenta correspond to stable, saddle-node unstable, AH unstable as well as fully unstable CW, respectively.
	The grey lines are cuts in the one-parameter planes $(\mu_\alpha,\,|E|^2)$ and $(Y_0,\,|E|^2)$ and are shown in panels (b) and (c), respectively. Parameters are  $(\delta_0,\,h,\,\eta,\,\varphi,\,\gamma,\,s)=(1,\,2,\,0.7,\,0,\,0.01,\,1)$.
}
	\label{fig:7}
\end{figure} 
\subsection{Bifurcation analysis}
To further investigate the stability regions of excitable TOPs and how they emerge, we perform a bifurcation analysis of the dynamical scenarios observed in Fig.~\ref{fig:6}.
First, setting $(\dot{E},\dot{\delta})=0$ in the system (\ref{eq:Eq4},\,\ref{eq:Eq5}), one can analytically calculate the emerging CW solutions and determine their linear stability, see Fig.~\ref{fig:7}. In panel (a) the intensity of the CW solution is plotted over the $(\mu_\alpha,\, Y_0)$ plane, whereas the corresponding cuts in $\mu_\alpha$ and $Y_0$-directions are shown in panels (b) and (c), respectively. The colors in Fig.~\ref{fig:7}~(a) represent the different types of instabilities, characterized by three discrete eigenvalues, see the bottom panel. In particular, a stable steady state (blue) can lose its stability in either a saddle-node bifurcation leading to a single real eigenvalue crossing the imaginary axis (orange) or a AH bifurcation (green), where a real part of the pair of complex eigenvalues changes its sign. Alternatively, all three eigenvalues can become unstable (magenta), however, this scenario is not realized for the parameters presented in Fig.~\ref{fig:7}.

We note that the AH unstable region can host both stable large amplitude relaxation oscillations and the narrow subthreshold oscillation regimes (cf. Fig.~\ref{fig:7}~(ii,iii)), whereas excitability regimes shown in Fig.~\ref{fig:7}~(i,iv) can be expected to occur close to the AH bifurcation line for negative values of $\mu_\alpha$. Indeed, in the focusing regime, where the strength of the Kerr effect $s$ is positive, hysteresis behavior can only be found if the $\delta$-nullcline is tilted into the same direction, which is the case for $\mu_\alpha<0$, cf. Fig.~\ref{fig:2}. In its turn, for negative $s$, one would get excitability for $\mu_\alpha>0$. Further, one can see  that for a large negative $\mu_\alpha<0$, the AH unstable regime seems to extend over an increasing $Y_0$ range, cf. Fig.~\ref{fig:7}. Since the bistability is lost at $\mu_\alpha=-1$, there are no dark excitable TOPs or dropouts to be expected beyond $\mu_\alpha=-1$.
\begin{figure}[t!]
	\centering
	\includegraphics[width=1\columnwidth]{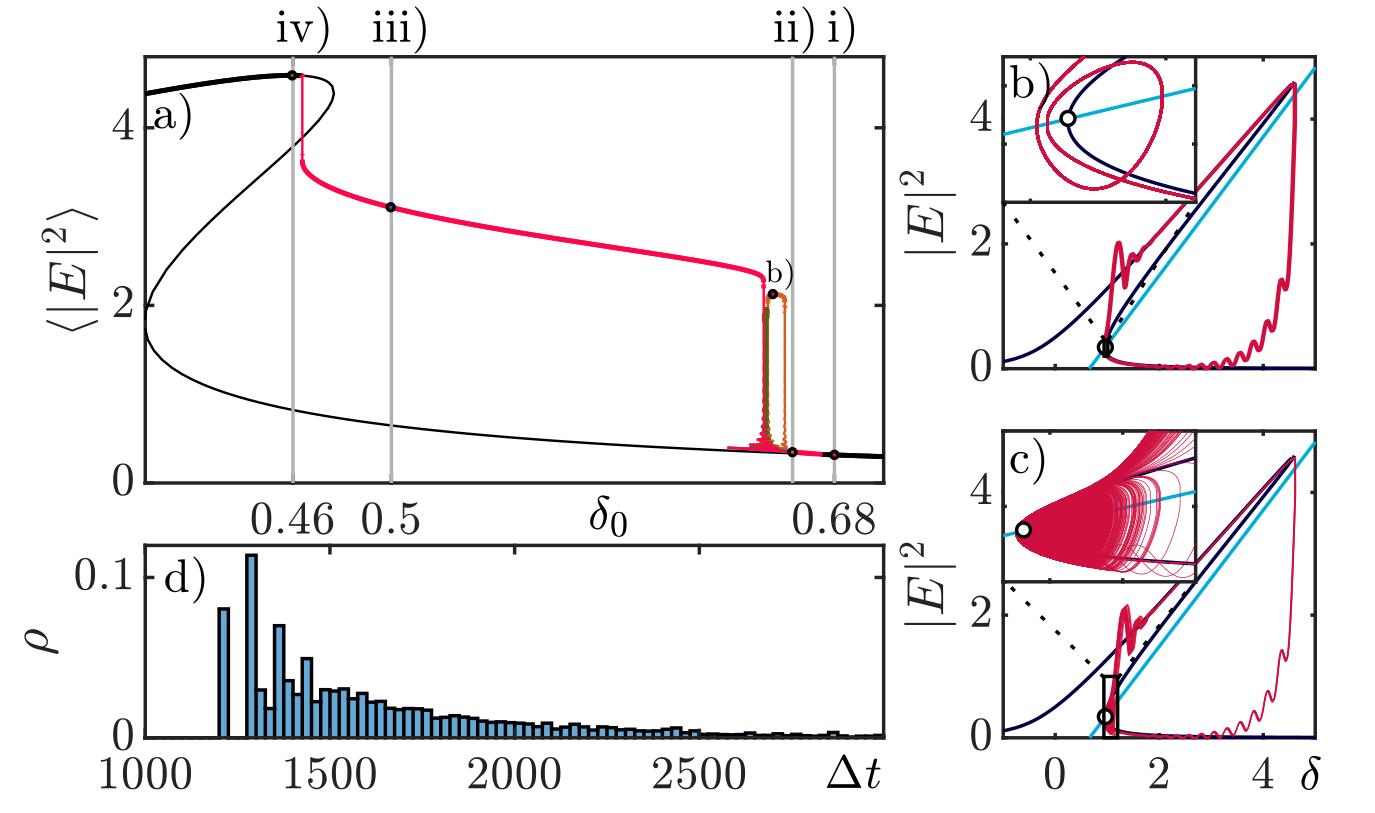}
	\caption{Bifurcation diagram obtained for $\gamma=0.05$ as a function of $\delta_0$. The branch of relaxation oscillations (red) emerges from the stable CW branch (solid black) in a supercritical AH bifurcation. Here, the subthreshold oscillation regime is very small, followed by a sharp Canard transition. After an extended relaxation-oscillation regime, another transition is entered, with a complex mixed-mode oscillation scenario. Grey vertical lines (i-iv) correspond to the dynamical regimes in Fig. \ref{fig:6}. b) Mixed mode oscillations with one subthreshold period. c) Deterministic random spiking due to a non-periodic attractor on the subthreshold branch surpassing the excitability threshold. Other parameters as in Fig.~\ref{fig:3}.}\label{fig:8}
\end{figure}
In the next step, we fix $\mu_\alpha$ and perform numerical path continuation. However, the small phase oscillations caused by the unbound phase dynamics in the fast stage of relaxation oscillations demand profiles with a highly refined mesh. Furthermore, continuation software often struggles to converge for extremely detailed solutions. For this reason, we reduce the scale separation, effectively decreasing the duration of relaxation oscillations by setting $\gamma=0.05$. Resulting branches are depicted in Fig.~\ref{fig:8} as a function of $\delta_0$. Here, vertical grey lines indicate corresponding dynamical regimes shown in panels (i)-(iv) in Fig.~\ref{fig:6}. One can see that for large values of $\delta_0$, the CW solution is stable (black thick line). However, decreasing $\delta_0$, a branch of periodic solutions (red) emerges close to point i) (bright excitable TOPs) in a supercritical AH bifurcation. Here, subthreshold oscillations (cf. point ii)) with a small period and amplitude emerge. Note, that applying noise on these periodic solutions will still produce bright excitable TOPs. Reducing $\delta_0$ further results in a canard explosion towards high oscillation amplitudes and periods leading to the formation of relaxation oscillations as at point iii). Further, another canard explosion at smaller values of $\delta_0$ results in a small region of subthreshold periodic oscillations around the upper CW branch, that finally becomes stable again in another supercritical AH bifurcation. Close to this point, dark excitable TOP can be formed in the presence of noise, see iv). Remarkably, as we approach the point along the vertical part of the periodic branch, where it passes the middle CW level, the period diverges, hinting an incomplete homoclinic as reported in \cite{AMC_NewJPhys_09} (cf. App.~\ref{appendixB} for more details).
\begin{figure}[t!]
	\centering
	\includegraphics[width=1\columnwidth]{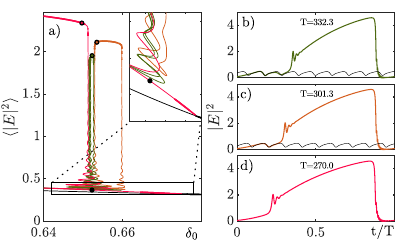}
	\caption{Zoom around the Canard explosion region found for $\gamma=0.05$ as a function of $\delta_0$.
	As the subthreshold oscillations (red) experience a period doubling bifurcation, they lose stability, while different mixed-mode oscillation branches (green, orange) emerge. (b-c) show exemplary profiles of mixed-mode oscillations at positions marked by filled circles featuring $n=2,1$ periods of the subthreshold oscillation. d) exemplary profile of the relaxation oscillation. Other parameters as in Fig.~\ref{fig:3}.}\label{fig:9}
\end{figure}

Figure \ref{fig:9}~(a) shows a zoom into the bifurcation diagram presented Fig.~\ref{fig:8} around the canard explosion region at large $\delta_0$. In this region, the branch of threshold oscillations (red) loses its stability in a period doubling bifurcation (cf. a blue branch in the inset of Fig.~\ref{fig:9}~(a)) and experiences a number of saddle-nodes bifurcations, finally resulting in a canard explosion. Here, additional two branches (green and orange) are shown that also feature stable oscillations with high amplitude and period. These solutions correspond to MMOs~\cite{SFM_PRE_05,MM_PRE_13, ZB_NonlinearDyn_15}, featuring an integer number of subthreshold oscillations followed by a large spike.

This behavior can be understood by looking at Fig.~\ref{fig:9} (b-d), where field intensity profiles over one period from each branch at the positions marked by filled circles (colored accordingly) are depicted. The black thin lines in panels (b-c) indicate the subthreshold oscillation solutions, taken from the $\delta_0$ value marked with a filled black circle that is visible in the insert in Fig.~\ref{fig:9}~(a). Figure~\ref{fig:9}~(d) shows the relaxation oscillations at the unstable CW solution. In contrary, in panel (c), a single period of the subthreshold oscillations appears additionally (cf. also Fig. \ref{fig:8}~(b) for the phase space dynamics), while in panel (b) two additional periods of of the subthreshold oscillation are observed. The period $T$ increases by $~30$ with each added subthreshold period which corresponds approximately to the period of the latter.

Note that all shown profiles in Fig.~\ref{fig:9}~(b-d) belong to stable parts of their respective branches and are thus accessible for numerical time simulations. This is shown in Fig.~\ref{fig:8}~(b), depicting a numerical solution for a $\delta_0$ value from the orange branch in Fig.~\ref{fig:9}~(c), indeed leading to the respective MMOs (cf. the inset of Fig.~\ref{fig:8}~(b)). In general MMOs can occur, when the subthreshold oscillations, for instance as result of a period-doubling bifurcation, surpass the threshold after an integer, in the case of chaotic spiking random, number of oscillation periods~\cite{MM_PRE_13}. In the present scenario, a cascade of period-doubling bifurcations, each introducing higher order MMOs, can result in a chaotic attractor on the subthreshold branch, surpassing the threshold in a non-periodic sequence. This allows for the observation of chaotic spiking in the absence of any noise as depicted in Fig.~\ref{fig:8}~(c). The respective ISI histogram in Fig.~\ref{fig:8}~(d) indicates the characteristic complex structure with multiple peaks resulting from the oscillatory structure of the chaotic attractor as  in~\cite{MMB_PRL_07}.

\section{Impact of the external cavity}
Our analysis demonstrated that a more realistic modeling including thermal effects leads to a wealth of new regimes.
Is is therefore justified to wonder if all the localized state solutions and their associated frequency combs can still be observed in this system. In particular, in the presence of the external cavity and in the absence of thermal effects (i.e. for $\mu_\alpha=0$), multistable bright and dark temporal localized states (TLSs) can be formed in the KGTI system~(\ref{eq:Eq1}-\ref{eq:Eq3}) in the normal dispersion regime via the locking of domain walls connecting the high and low intensity levels of the injected micro-cavity~\cite{SPV-OL-19,SJG-OL-22}. Note that as even if the external cavity can be as long as several centimeters, the thermal timescale remains much longer than the cavity round-trip. Hence, in the following we focus on the regime where $\gamma^{-1}\gg\tau>1$, setting time-delayed feedback on a time scale between the slow thermal effects and the fast micro-cavity dynamics. Further, we are interested in the impact of the external cavity on the formation of excitable TOPs.
\begin{figure}[t!]
	\centering
	\includegraphics[width=1\columnwidth]{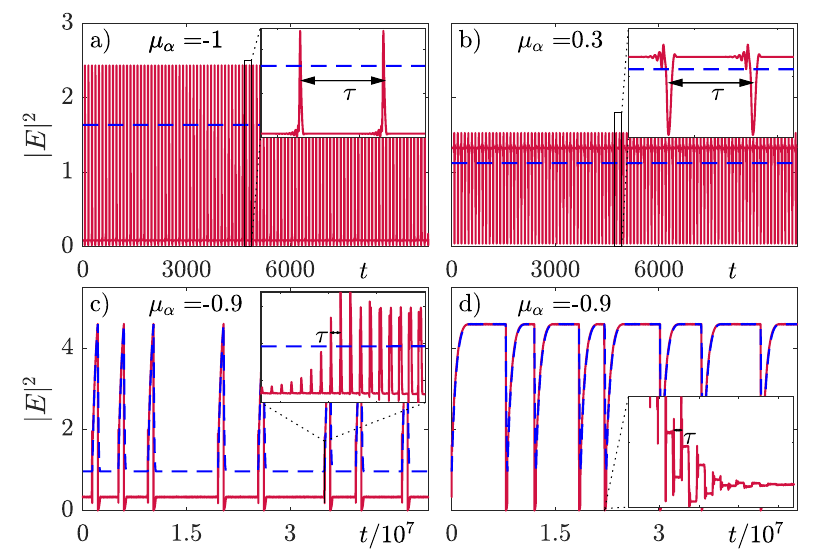}
	\caption{Time simulations of Eqs.~(\ref{eq:Eq1}-\ref{eq:Eq3}) for $\tau=100$. (a,b) Stable bright and dark TLSs for $\gamma=10^{-4}$ and $(\delta_0,h,\eta,\varphi,Y_0)=(1.5,2,0.75,0,0.6/0.38)$.
	 (c,d) Thermo-optical excitability observed for $\gamma=10^{-5}$ and other parameters as in Fig. \ref{fig:3} (Fig. \ref{fig:4}).}\label{fig:10}
\end{figure}
We fix $\mu_\alpha$ to a value yielding one stable fixed point (cf. Fig.~\ref{fig:2} (b) for $\mu_\alpha=-1$) and perform numerical simulations of Eqs.~(\ref{eq:Eq1}-\ref{eq:Eq3}) in the normal dispersion regime.
Our results are presented in Fig.~\ref{fig:10}~(a), where the $\tau$-periodic train of bright TLS is shown in red, while the blue dashed line corresponds to the time evolution of the thermal detuning $\delta$. Remarkably, in this regime, the thermal effects allow TLSs to remain stable outside of the bistability region since for $\mu_\alpha=-1$ the bistability is lost (cf. Fig.~\ref{fig:2} (b)). The absence of the bistability changes the nature of the resulting TLS as they were previously born from locking of fronts between two stable CW values, whereas now they are homoclinic orbits living on a monostable CW background (cf. the inset in Fig.~\ref{fig:10}~(a)). Consequently, the dark TLSs observed for the same parameter set~\cite{SJG-OL-22} cannot be stabilized for $\mu_\alpha=-1$ as they live on the, now non-existing, upper CW background. However, they can be found for smaller values of $\mu_\alpha$ as shown in Fig.~\ref{fig:10}~(b) for $\mu_\alpha=0.3$.

Finally, bright and dark TOP excitations can be observed for the same  parameter set as in Fig.~\ref{fig:6} in the presence of the external cavity, see Fig.~ \ref{fig:10}~(c,d), respectively. Here, a new phenomenon can be observed, as the small oscillations caused by the phase dynamics are now oscillating with a period close to the delay time $\tau$. In particular, in the inset of Fig.~\ref{fig:10}~(c), these oscillations even resemble TLSs that might occur temporarily due to the slow $\delta$ dynamics as they pass certain parameter values. The occurrence of both timescales in the same orbit clearly indicates how these timescales are able to interact.

However, the excitability seems to vanish if $\delta$ does not evolve on the slowest timescales. This observation is supported if $\gamma$ is increased to e.g., $\gamma=10^{-4}$, leading to a situation where the excitability competes with a $\tau$ periodic solution (cf. Fig.~ \ref{fig:10}~(a,b)). Considering another limit case, where $\tau\gg \gamma^{-1}$, excitable orbits would turn into periodic solutions, as the excitation would travel through the external cavity while the thermal dynamics would fully relax. As a result, the optical feedback would excite the system repeatably after each round-trip. However, this scenario is of no experimental use for this specific setup.

\section{Conclusion}
We investigated the influence of thermal effects on the dynamics of a vertically emitting micro-cavity with a Kerr nonlinearity, subjected to detuned optical injection and coupled to a long external cavity. As the microcavity heats in the presence of light, the reflective index and with that the micro-cavity resonance evolve on a slow time scale. Consequently, the detuning between injection frequency and micro-cavity resonance turns into a dynamical variable. To isolate the influence of thermal effects, we started with a vanishing delay case, reducing the system to a three-dimensional ordinary differential equation. Here, we find excitability close to a supercritical canard scenario. In the vicinity of a bistable continuous wave (CW) regime caused by the Kerr nonlinearity, we observed dark and bright excitability on the higher and lower CW background, respectively. In the bright case, mixed-mode oscillations and even chaotic spiking were demonstrated. Finally, we added an external cavity with a delay that is long compared to the neglected microcavity delay, but short compared to the thermal timescale. We confirmed, that localized states observed in previous works continue to exist, if thermal effects are included, while the new relaxation oscillation regimes create stable localized states beyond the CW bistability. This shall act as an outlook on future investigations regarding the interaction between delay and thermal effects. Finally, our extended model is a first step towards elaborating strategies regarding thermal cavity stabilization and cavity cooling using, e.g., an additional control beam and a secondary resonance as detailed in ~\cite{DSBP_NatPhotonics_20}.


\begin{acknowledgments}
$\,$
\small
\textbf{Funding} 
J.J. acknowledges the financial support of the project KEFIR/AEI/10.13039/501100011033/ FEDER, UE. E.R.K, J.J. and S.V.G. acknowledge the financial support of the project KOGIT, Agence Nationale de la Recherche (ANR-22-CE92-0009)), Deutsche Forschungsgemeinschaft (DFG) via Grant Nr. 505936983.
$\,$

\textbf{Disclosures} The authors declare no conflicts of interest
$\,$

\textbf{Data availability} Data underlying the results presented in this paper are not publicly 
available  at this time but may be obtained from the authors upon reasonable request.
$\,$
\end{acknowledgments}

\appendix

\section{Model derivation}
\label{appendixA}

We derive in this section the Kerr Gires-Tournois interferometer model
from first principles giving particular attention to how thermal effects
are included. We start from the Maxwell equations for the real electric
and magnetic fields so that $\left(\boldsymbol{\mathcal{E}},\boldsymbol{\mathcal{B}}\right)\in\mathbb{R}^{3}$.
We assume a nonmagnetic material as well as the absence of free charges
and free currents, yielding 
\begin{align}
\nabla\cdot\boldsymbol{\mathcal{D}} & =0,\label{eq:GE}\\
\nabla\cdot\boldsymbol{\mathcal{B}} & =0,\label{eq:GM}\\
-\nabla\times\boldsymbol{\mathcal{E}} & =\dot{\boldsymbol{\mathcal{B}}},\label{eq:MF}\\
\nabla\times\boldsymbol{\mathcal{H}} & =\dot{\boldsymbol{\mathcal{D}}}.\label{eq:MA}
\end{align}
The auxiliary fields $\left(\boldsymbol{\mathcal{D}},\boldsymbol{\mathcal{H}}\right)$
are defined as 
\begin{align}
\boldsymbol{\mathcal{D}} & =\epsilon_{0}\boldsymbol{\mathcal{E}}+\boldsymbol{\mathcal{P}}\label{eq:AuxD}\\
\boldsymbol{\mathcal{H}} & =\boldsymbol{\mathcal{B}}/\mu_{0},\label{eq:AuxH}
\end{align}
where $\boldsymbol{\mathcal{P}}=\boldsymbol{\mathcal{P}}_{\mathrm{l}}+\boldsymbol{\mathcal{P}}_{\mathrm{nl}}$
is the polarization, containing both the linear response of the cavity,
in particular the refractive index dependence on the temperature,
as well as the Kerr nonlinearity. For an isotropic and homogeneous
medium, the constitutive relation for the refractive index is given
by a convolution $\boldsymbol{\mathcal{P}}_{\mathrm{l}}=\varepsilon_{0}\chi_{l}\star\boldsymbol{\mathcal{E}}$,
i.e.
\begin{eqnarray}
\boldsymbol{\mathcal{P}}_{\mathrm{l}}\left(\mathbf{r},t\right) & = & \frac{\epsilon_{0}}{2\pi}\int_{\mathbb{R}}\chi_{l}\left(s\right)\boldsymbol{\mathcal{E}}\left(\mathbf{r},t-s\right)ds\label{eq:Conv}
\end{eqnarray}
where $\chi_{l}\left(s\right)$ is a causal function. Assuming that
the nonlinear effects are small, i.e. $\boldsymbol{\mathcal{P}}_{\mathrm{nl}}\ll\boldsymbol{\mathcal{P}}_{\mathrm{l}}$
we have that $\nabla\cdot\boldsymbol{\mathcal{P}}\simeq\nabla\cdot\boldsymbol{\mathcal{P}}_{\mathrm{l}}$.
Taking the divergence of Eq.~\ref{eq:AuxD}, using Eq.~\ref{eq:GE}
and noting that Eq.~\ref{eq:Conv} is an integral in the time domain,
we deduce that both $\boldsymbol{\mathcal{E}}$ and $\boldsymbol{\mathcal{P}}_{\mathrm{l}}$
are divergence-free, i.e. $\nabla\cdot\boldsymbol{\mathcal{E}}\simeq0$
and $\nabla\cdot\boldsymbol{\mathcal{P}}_{\mathrm{l}}\simeq0$.

In order to describe the propagation of electromagnetic waves we applying
the curl operator to Maxwell-Faraday equation given in Eq.~\ref{eq:MF}.
Using Maxwell-Ampere equation given by Eq.~\ref{eq:MA}, the result
simplifies into 
\begin{align}
-\mu_{0}\ddot{\boldsymbol{\mathcal{D}}} & =\nabla\times\left(\nabla\times\boldsymbol{\boldsymbol{\mathcal{E}}}\right).
\end{align}

Using the vector identity $\nabla\times\left(\nabla\times\boldsymbol{\boldsymbol{\mathcal{E}}}\right)=\nabla\left(\nabla\cdot\boldsymbol{\mathcal{E}}\right)-\nabla^{2}\boldsymbol{\mathcal{E}}$
and that $\boldsymbol{\boldsymbol{\mathcal{E}}}$ is approximately
divergence-free we get 
\begin{eqnarray}
\nabla^{2}\boldsymbol{\mathcal{E}}-\epsilon_{0}\mu_{0}\partial_{t}^{2}\boldsymbol{\mathcal{E}} & = & \mu_{0}\partial_{t}^{2}\left(\boldsymbol{\mathcal{P}}_{\mathrm{l}}+\boldsymbol{\mathcal{P}}_{\mathrm{nl}}\right).\label{eq:WaveEq}
\end{eqnarray}

In the rest of the work, we shall consider a quasi-monochromatic field
around the carrier angular frequency of the optical injection $\omega_{0}$.
We further assume that all the fields are transversely polarized,
say in the $y$-direction, so that $\boldsymbol{\mathcal{E}}\left(\mathbf{r},t\right)=\mathcal{E}\left(\mathbf{r},t\right)\boldsymbol{\mathbf{y}}$
with $\mathcal{E}$ a real scalar amplitude. We define the complex
field envelope $E\left(\mathbf{r},t\right)$ that is slowly evolving
in time such that
\begin{eqnarray}
\mathcal{E}\left(\mathbf{r},t\right) & = & \frac{1}{2}\left[E\left(\mathbf{r},t\right)e^{-i\omega_{0}t}+\mathrm{c.c.}\right]\label{eq:fieldansatz}
\end{eqnarray}

Similarly, the linear and nonlinear polarizations can be expressed
as $\boldsymbol{\mathcal{P}}_{\mathrm{l,nl}}\left(\mathbf{r},t\right)=\mathcal{P}_{\mathrm{l,nl}}\left(\mathbf{r},t\right)\mathbf{y}$
with 
\begin{eqnarray}
\mathcal{P}_{\mathrm{l}}\left(\mathbf{r},t\right) & = & \frac{1}{2}\left[P_{\mathrm{l}}\left(\mathbf{r},t\right)e^{-i\omega_{0}t}+\mathrm{c.c.}\right]
\end{eqnarray}
where the expression of $P_{\mathrm{l}}$ reads after using Eq.~\ref{eq:Conv}
\begin{eqnarray}
P_{\mathrm{l}}\left(\mathbf{r},t\right) & = & \frac{1}{2\pi}\int_{\mathbb{R}}\epsilon_{0}\chi_{l}\left(s\right)E\left(\mathbf{r},t-s\right)e^{i\omega_{0}s}ds.\label{eq:Plin}
\end{eqnarray}

We define the Kerr nonlinearity as $\boldsymbol{\mathcal{P}}_{\mathrm{nl}}=\varepsilon_{0}\chi_{3}\left(\mathbf{r}\right)\mathcal{E}^{3}\boldsymbol{y}$
and we assume the Kerr medium to be located over a finite region within
the cavity, which is achieved by considering that $\chi_{3}$ is a
function of $\mathbf{r}$. We define similarly
\begin{eqnarray}
\mathcal{P}_{\mathrm{nl}}\left(\mathbf{r},t\right) & = & \frac{1}{2}\left[P_{\mathrm{nl}}\left(\mathbf{r},t\right)e^{-i\omega_{0}t}+\mathrm{c.c.}\right]+\mathrm{h.o.t}\label{eq:Pnlin}
\end{eqnarray}
where $\mathrm{h.o.t}$ denotes non-resonant terms at frequencies
$\pm3\omega_{0}$ while the resonant contribution $P_{\mathrm{nl}}$
reads
\begin{eqnarray}
P_{\mathrm{nl}}\left(\mathbf{r},t\right) & = & \frac{3}{4}\varepsilon_{0}\chi_{3}\left(\mathbf{r}\right)\left|E\right|^{2}E\left(\mathbf{r},t\right)
\end{eqnarray}

The Fourier transform operator $\mathcal{F}_{t}$ is defined as
\begin{eqnarray}
\mathcal{F}_{t}\left[\mathcal{A}\right] & = & \frac{1}{2\pi}\int_{\mathbb{R}}\mathcal{A}\left(\mathbf{r},t\right)e^{i\omega t}dt,
\end{eqnarray}
while the inverse transform $\mathcal{F}_{t}^{-1}$ reads 
\begin{eqnarray}
\mathcal{F}_{t}^{-1}\left[\mathcal{A}\right] & = & \int_{\mathbb{R}}\mathcal{A}\left(\mathbf{r},\omega\right)e^{-i\omega t}d\omega.
\end{eqnarray}

For the sake of simplicity, we use identical letters for the function
in the time and in the frequency domain. Finally, the Fourier transform
of $\boldsymbol{\mathcal{E}}$ can be expressed as a function
of $E$ since 
\begin{eqnarray}
\mathcal{E}\left(\mathbf{r},\omega\right) & = & \frac{1}{2}\left[E\left(\mathbf{r},\omega-\omega_{0}\right)+E^{\star}\left(\mathbf{r},-\omega-\omega_{0}\right)\right]
\end{eqnarray}

\subsection{micro-cavity with a thin nonlinear region}

As light propagates along the cavity axis $z$ we denote by $\mathbf{r}_{\perp}=\left(x,y\right)$
the transverse coordinates so that $E\left(\mathbf{r},t\right)=E\left(\mathbf{r}_{\perp},z,t\right)$
while $\Delta=\partial_{z}^{2}+\Delta_{\perp}^{2}$. Considering only
the positive frequencies in Eq.~\ref{eq:WaveEq} we find in the temporal
Fourier domain that
\begin{eqnarray}
\left(\partial_{z}^{2}+\Delta_{\perp}^{2}+\frac{\omega^{2}}{\upsilon^{2}}\right)E\left(\mathbf{r}_{\perp},z,\omega-\omega_{0}\right) & = & -\mu_{0}\omega^{2}P_{\mathrm{nl}}\label{eq:WaveEq_pw}
\end{eqnarray}
where we defined the frequency dependent velocity $\upsilon^{2}=\left[c/n\left(\omega\right)\right]^{2}$
and the frequency dependent refractive index $n\left(\omega\right)=\sqrt{1+\chi\left(\omega\right)}$.
We consider a thin Kerr slice of width $W$ located in $z=l$ that
we model as $\chi_{3}\left(\mathbf{r}\right)=\bar{\chi}_{3}W\delta\left(z-l\right)$.
Taking the transverse Fourier transform of Eq.~\ref{eq:WaveEq_pw}
so that $\Delta_{\perp}^{2}\rightarrow-q_{\perp}^{2}$ and integrating
around $z=l$ over a range $2\varepsilon$ leads to\begin{widetext}
\begin{eqnarray}
\left.\partial_{z}E\left(\mathbf{q}_{\perp},z,\omega-\omega_{0}\right)\right|_{l-\varepsilon}^{l+\varepsilon}+\int_{l-\varepsilon}^{l+\varepsilon}\left(\frac{\omega^{2}}{\upsilon^{2}}-q_{\perp}^{2}\right)E\left(\mathbf{q}_{\perp},z',\omega-\omega_{0}\right)dz' & = & -\frac{3\omega^{2}\bar{\chi}_{3}W}{4c^{2}}\mathcal{F}_{t,\perp}\left[\left|E\right|^{2}E\left(\mathbf{r}_{\perp},l,\omega-\omega_{0}\right)\right].\label{eq:WaveEq_pwq}
\end{eqnarray}
\end{widetext}

The next step of our derivation consists in splitting the electromagnetic
field in each part of the cavity before and after the Kerr slice.
For a micro-cavity of length $L$ with a nonlinear medium placed at
$z=l$ we define 
\begin{align}
E(\mathbf{q}_{\perp},z,\omega)= & \begin{cases}
L\left(\mathbf{q}_{\perp},z,\omega\right) & 0<z<l\\
R\left(\mathbf{q}_{\perp},z,\omega\right) & l<z<L,
\end{cases}
\end{align}

The field in each sub-section is decomposed into counter-propagating
modes along the cavity axis
\begin{eqnarray}
L\left(\mathbf{q}_{\perp},z,\omega\right) & = & L_{+}\left(\mathbf{q}_{\perp},\omega\right)e^{iqz}+L_{-}\left(\mathbf{q}_{\perp},\omega\right)e^{-iqz}\label{eq:Lpm}\\
R\left(\mathbf{q}_{\perp},z,\omega\right) & = & R_{+}\left(\mathbf{q}_{\perp},\omega\right)e^{iqz}+R_{-}\left(\mathbf{q}_{\perp},\omega\right)e^{-iqz}\label{eq:Rpm}
\end{eqnarray}

We defined in Eqs.~(\ref{eq:Lpm},\ref{eq:Rpm}) the frequency-dependent
longitudinal wave vector 
\begin{eqnarray}
q(\mathbf{q}_{\perp},\omega) & = & \sqrt{\omega^{2}/\upsilon^{2}-q_{\perp}^{2}}\label{eq:DispRel}
\end{eqnarray}
The continuity of the transverse field at the boundary in $z=l$ imposes
that 
\begin{eqnarray}
L\left(\mathbf{q}_{\perp},l,\omega\right) & = & R\left(\mathbf{q}_{\perp},l,\omega\right)=\mathtt{E}\left(\mathbf{q}_{\perp},\omega\right)
\end{eqnarray}
where we defined $\mathtt{E}\left(\mathbf{q}_{\perp},\omega\right)$
as the field amplitude impinging on the Kerr slice in $z=l$. The
respective boundary condition for the derivative discontinuity imposed
by the nonlinear medium taking the limit $\varepsilon\rightarrow0$
in Eq.~\ref{eq:WaveEq_pwq} reads
\begin{eqnarray}
\partial_{z}R-\partial_{z}L & = & -\frac{3\omega^{2}\bar{\chi}_{3}W}{4c^{2}}\mathcal{F}_{t,\perp}\left[\left|\mathtt{E}\right|^{2}\mathtt{E}\right]\label{eq:Discont}
\end{eqnarray}
where we omitted the dependence of all the variables on $\left(\mathbf{q}_{\perp},l,\omega-\omega_{0}\right)$
for clarity.

The additional boundary conditions linking the micro-cavity with the
exterior can be deduced from the setup depicted in Fig.~1. We evaluate
the fields at the two extreme sides of the micro-cavity ($z=0,L$)
for which the two distributed Bragg mirrors have reflectivities in
the frequency domain $r_{1,2}(\omega)$ and $t_{1,2}(\omega)$. The
primed indexes are for transmission and reflection processes that
originate from outside of the cavity. We assume that the DBRs have
a wide bandwidth and, as such, that the modulus of the reflection
coefficients do not depends on the frequency, i.e. $r_{1,2}\left(\omega\right)=\left|r_{1,2}\right|\exp\left[i\phi_{j}\left(\omega\right)\right]$.
The output of the micro-cavity is a superposition of the reflection
of the injected field as well as the micro-cavity transmission of
its internal field and reads 
\begin{eqnarray}
O & = & t_{1}L_{-}+r_{1}^{\prime}Y\label{eq:Out}
\end{eqnarray}
 while the intra-cavity counter-propagating waves are linked by the
mirror reflectivies
\begin{eqnarray}
L_{+}=r_{1}L_{-}+t_{1}^{\prime}Y & \;,\; & R_{-}e^{-iqL}=r_{2}R_{+}e^{iqL}\label{eq:BC_LR}
\end{eqnarray}
Combining Eqs.~(\ref{eq:Discont},\ref{eq:BC_LR}) allows to express
$\left(L_{\pm},R_{\pm}\right)$ as functions of the field on the Kerr
medium $\mathtt{E}\left(\mathbf{q}_{\perp},\omega\right)$ and the
injected field
\begin{eqnarray}
R_{+}=\frac{e^{-iql}\mathtt{E}}{1+r_{2}e^{2iq(L-l)}} & \;,\; & R_{-}=\frac{r_{2}e^{iql}\mathtt{E}}{r_{2}+e^{2iq(l-L)}},\label{eq:RpmE}\\
L_{+}=\frac{r_{1}e^{iql}\mathtt{E}+t_{1}^{\prime}Y}{1+r_{1}e^{2iql}} & \;,\; & L_{-}=\frac{e^{-iql}\mathtt{E}-t_{1}^{\prime}Y}{r_{1}+e^{-2iql}}.\label{eq:LpmE}
\end{eqnarray}

These expressions can now be inserted into the derivative boundary
condition Eq.~\ref{eq:Discont} which yields 
\begin{eqnarray}
\left(R_{+}-L_{+}\right)e^{iql}-\left(R_{-}-L_{-}\right)e^{-iql} & =\label{eq:Discont-1}\\
-\frac{3\omega^{2}\bar{\chi}_{3}W}{4iqc^{2}}\mathcal{F}_{t,\perp}\left[\left|\mathtt{E}\right|^{2}\mathtt{E}\right]\nonumber 
\end{eqnarray}

Upon simplification using Eqs.~(\ref{eq:RpmE},\ref{eq:LpmE}) as
well as the Stokes relations $\left(r=-r',tt'-rr'=1\right)$ we deduce
in the transverse and temporal Fourier domains the fundamental equation
linking the value of the intra-cavity field on the Kerr slice $\mathtt{E}$
with that of the injected field impinging upon the micro-cavity $Y$
\begin{eqnarray}
\negthickspace\negthickspace\negthickspace\negthickspace F_{1}\mathtt{E} & = & -\frac{3\omega^{2}\bar{\chi}_{3}W}{8iqc^{2}}\Gamma\mathcal{F}_{t,\perp}\left[\left|\mathtt{E}\right|^{2}\mathtt{E}\right]+F_{2}Y.\label{eq:MB}
\end{eqnarray}
where we introduced the confinement factor $\Gamma$ as well as the
functions $F_{1,2}$ that are characteristic of the Fabry-Perot resonator
\begin{eqnarray}
\Gamma\left(q,\omega\right) & = & \left(1+r_{1}e^{2iql}\right)\left(1+r_{2}e^{2iq(L-l)}\right)\label{eq:Gma}\\
F_{1}\left(q,\omega\right) & = & 1-r_{1}r_{2}e^{2iqL},\label{eq:F1}\\
F_{2}\left(q,\omega\right) & = & t_{1}^{\prime}e^{iql}\left(1+r_{2}e^{2iq(L-l)}\right).\label{eq:F2}
\end{eqnarray}

The modes of the micro-cavity correspond to the minima of $\left|F_{1}\left(q,\omega\right)\right|$.
They define the cavity resonances for which the intra-cavity field
is maximum. This condition is achieved when 
\begin{eqnarray}
2q_{m}L & = & 2\pi m-\left[\phi_{1}\left(\omega_{m}\right)+\phi_{2}\left(\omega_{m}\right)\right]\label{eq:ModeFP}
\end{eqnarray}
where $m\in\mathrm{N}$. On another hand, the Kerr slice is placed
in order to maximize the confinement factor under normal incidence,
i.e. at a field anti-node. This is achieved by imposing that
\begin{eqnarray}
2q_{n_{1},n_{2}}l+\phi_{1}\left(\omega_{n_{1},n_{2}}\right) & = & 2\pi n_{1}\label{eq:GmaMax1}\\
2q_{n_{1},n_{2}}\left(L-l\right)+\phi_{2}\left(\omega_{n_{1},n_{2}}\right) & = & 2\pi n_{2}\label{eq:GmaMax2}
\end{eqnarray}
where $n_{1,2}\in\mathrm{N}$. Notice that summing Eqs.~(\ref{eq:GmaMax1},\ref{eq:GmaMax2})
leads again to Eqs.~\ref{eq:ModeFP} with $m=n_{1}+n_{2}$. As such,
solving Eqs.~(\ref{eq:GmaMax1},\ref{eq:GmaMax2}) allows finding
all of the micro-cavity modes. The most sensible choice corresponds
to placing the Kerr medium at the center of the cavity, i.e. $l=\frac{L}{2}$.
In this situation, we find that one every other modes of the cavity
maximizes $\Gamma$. We select the cavity mode $q_{m}=2\pi m/L$ with
$m\in\mathbb{N}$ for which the associated frequency $\omega_{m}$,
as defined by the dispersion relation $q_{m}=\omega_{m}/\upsilon\left(\omega_{m}\right)$,
is the closest to the injection frequency $\omega_{0}$ and we expand
$F_{1}$, $F_{2}$ and $\Gamma$ around the mode $\left(q_{m},\omega_{m}\right)$.
In the good cavity limit for which $\left|r_{1,2}\right|\rightarrow1$,
the resonances are sharp and $\Gamma\rightarrow4$ and $\left|F_{2}\right|\rightarrow2t_{1}^{\prime}$
vary much more slowly than $F_{1}$. Noticing further that $q_{m}\sim L^{-1}$,
we find the following hierarchy regarding the relative variation of
$F_{1,2}$ and $\Gamma$
\begin{eqnarray}
L^{-1}F_{1}^{-1}\frac{dF_{1}}{dq}\left(q_{m}\right) & \sim & \frac{1}{1-\left|r_{1}r_{2}\right|}\gg1\\
L^{-1}F_{2}^{-1}\frac{dF_{2}}{dq}\left(q_{m}\right) & \sim & 1,\\
L^{-1}\left(\frac{\Gamma}{q}\right)^{-1}\frac{d}{dq}\left(\frac{\Gamma}{q}\right)\left(q_{m}\right) & \text{\ensuremath{\sim}} & 1.
\end{eqnarray}

It is therefore sensible to consider only the lowest order for $F_{2}$
and $\Gamma$. However, in order to back-transform Eq.~\ref{eq:ModeFP}
to the spatiotemporal domain $F_{1}$ needs to be expanded to first
order in $\left(\omega,q_{\perp}\right)$. We consider small frequency
variations and we assume $\omega$ to be close to a cavity mode. As
the temperature varies, the refractive index $n(\omega,T)$ changes,
effectively shifting the cavity length and hence the resonance frequency
positions.
We note that in many semiconductors such as InP, SiC or GaAs, the thermo-optic effect $\partial_T n$ is 
often an order of magnitude larger than the effect of thermal expansion $L^{-1}\partial_T L$.
For example in GaAs $L^{-1}\partial_T L \sim 6 \times 10^{-6}\,\mathrm{K}^{-1}$
while $\partial_T n \sim 2.35 \times 10^{-4}\,\mathrm{K}^{-1}$\cite{DCI-APL-00}.
For these reasons, we shall neglect in what follows cavity length variations 
due to thermal effects and we consider solely the thermo-optic effect.
We allow for small temperature deviations around the equilibrium
temperature $T_{0}$ and we define $\delta T=T-T_{0}\ll1$. We define
$\bar{\omega}_{m}$ as the mode frequency when the cavity is at temperature
$T_{0}$ and introduce the small deviation $\delta\omega=\omega-\bar{\omega}_{m}\ll1$.
The reference refractive index is $\bar{n}_{m}=n\left(\bar{\omega}_{m},T_{0}\right)$
and we set $\bar{\upsilon}_{m}=c/\bar{n}_{m}$ while $\bar{q}_{m}=\bar{\omega}_{m}/\bar{\upsilon}_{m}$
solves Eq.~\ref{eq:ModeFP}. Finally, we allow for shallow transverse
dynamics an assume that $q_{\perp}\ll1$. To first order in the deviations
Eq.~\ref{eq:DispRel} reads 
\begin{eqnarray}
q & \simeq & \bar{q}_{m}+\frac{\delta\omega}{\upsilon_{e}}+\bar{q}_{m}\frac{\partial_{T}\bar{n}_{m}}{\bar{n}_{m}}\delta T-\frac{q_{\perp}^{2}}{2\bar{q}_{m}},\label{eq:DisPara}
\end{eqnarray}
where we defined the effective group index velocity $\upsilon_{e}$
taking into account the dispersion of the refractive index
\begin{eqnarray}
\frac{1}{\upsilon_{e}} & = & \frac{1}{\bar{\upsilon}_{m}}+\bar{q}_{m}\frac{\partial_{\omega}\bar{n}_{m}}{\bar{n}_{m}}.
\end{eqnarray}

Inserting Eq.~\ref{eq:DisPara} in Eq.~\ref{eq:MB} and denoting
$\Delta n=\partial_{T}\bar{n}_{m}\delta T$ yields \begin{widetext} 

\begin{equation}
\mathtt{E}\left[F_{1}(\bar{q}_{m},\bar{\omega}_{m})+\partial_{q}F_{1}(\bar{q}_{m},\bar{\omega}_{m})\left(\bar{q}_{m}\frac{\Delta n}{\bar{n}_{m}}-\frac{q_{\perp}^{2}}{2\bar{q}_{m}}\right)-i\delta\omega\tau_{e}\left|r_{1}r_{2}\right|\right]=-\frac{3\bar{\omega}_{m}\bar{\chi}_{3}W}{8i\bar{n}_{m}c}\Gamma\left(\bar{q}_{m},\bar{\omega}_{m}\right)\mathcal{F}_{t,\perp}\left[\left|\mathtt{E}\right|^{2}\mathtt{E}\right]+F_{2}(\bar{q}_{m},\bar{\omega}_{m})Y,
\end{equation}
\end{widetext}  with $F_{1}(\bar{q}_{m},\bar{\omega}_{m})=1-\left|r_{1}r_{2}\right|$
and $\partial_{q}F_{1}(\bar{q}_{m},\bar{\omega}_{m})=-2i\left|r_{1}r_{2}\right|L$.
We introduced the cavity round-trip $\tau_{e}=2L_{e}/\upsilon_{e}$
and the effective cavity length $L_{e}$. The latter depends, beyond
the standard contribution of the cavity length, upon the penetration
length in the DBRs, i.e. $L_{e}=L+\upsilon_{e}\partial_{\omega}\left[\left(\phi_{1}+\phi_{2}\right)/2\right]\left(\bar{\omega}_{m}\right)$
while at zeroth order we have
\begin{eqnarray}
F_{2}(\bar{q}_{m},\bar{\omega}_{m}) & \simeq & t_{1}^{\prime}\left(-1\right)^{n_{1}}e^{-i\phi_{1}\left(\bar{\omega}_{m}\right)/2}(1+\left|r_{2}\right|)\\
\Gamma(\bar{q}_{m},\bar{\omega}_{m}) & \simeq & (1+\left|r_{1}\right|)(1+\left|r_{2}\right|).
\end{eqnarray}

The photon decay rate in amplitude is $\kappa=\tau_{e}^{-1}\left(1-\left|r_{1}r_{2}\right|\right)/\left|r_{1}r_{2}\right|$,
and we define the coupling parameter $\tilde{h}=F_{2}(\bar{q}_{m},\bar{\omega}_{m})/F_{1}(\bar{q}_{m},\bar{\omega}_{m})$,
which leads us to 
\begin{eqnarray}
\left[1-\frac{i\delta\omega}{\kappa}-\frac{2iL\left|r_{1}r_{2}\right|}{1-\left|r_{1}r_{2}\right|}\left(\bar{q}_{m}\frac{\Delta n}{n_{m}}-\frac{q_{\perp}^{2}}{2\bar{q}_{m}}\right)\right]\mathtt{E} & = & \tilde{h}Y\nonumber \\
-\frac{3\bar{\omega}_{m}\bar{\chi}_{3}W}{8i\bar{n}_{m}c}\frac{(1+\left|r_{1}\right|)(1+\left|r_{2}\right|)}{1-\left|r_{1}r_{2}\right|}\mathcal{F}_{t,\perp}\left[\left|\mathtt{E}\right|^{2}\mathtt{E}\right]
\end{eqnarray}

By back-transforming to time and transverse space, noting that $\delta\omega=\omega-\omega_{0}+\omega_{0}-\omega_{m}$
and that the field is is a function of $\omega-\omega_{0}$ we get
\begin{eqnarray}
\frac{1}{\kappa}\frac{\partial\mathtt{E}}{\partial t} & = & -\left(1+i\delta_{0}\right)\mathtt{E}+\tilde{h}Y+\frac{i2L\left|r_{1}r_{2}\right|}{1-\left|r_{1}r_{2}\right|}\left(\bar{q}_{m}\frac{\Delta n}{n_{m}}+\frac{\Delta_{\perp}}{2\bar{q}_{m}}\right)\mathtt{E}\nonumber \\
 & + & \frac{3i\bar{\omega}_{m}\bar{\chi}_{3}W}{8\bar{n}_{m}c}\frac{(1+\left|r_{1}\right|)(1+\left|r_{2}\right|)}{1-\left|r_{1}r_{2}\right|}|\mathtt{E}|^{2}\mathtt{E}\label{eq:Bordel}
\end{eqnarray}
where we defined $\delta_{0}=\omega_{m}-\omega_{0}$ the detuning
at temperature $T_{0}$ between the cavity mode and the injection. 

The evolution equation for the intra-cavity field given by Eq.~\ref{eq:Bordel}
is unwieldly as it depends on the DBR transmission $t_{1}^{\prime}$
trough the parameter $\tilde{h}$. However, this dependence can be
simplified by a proper scaling of $\mathtt{E}$. Starting from the
output relation Eq.~\ref{eq:Out} and the expression of $L_{-}$
given by Eq.~\ref{eq:LpmE} we observe, after using the Stokes relations
and $l=L/2$ that $O=\alpha\mathtt{E}-Y$ with $\alpha=\left(-1\right)^{n_{1}}e^{-i\phi_{1}\left(\bar{\omega}_{m}\right)/2}t_{1}/\left(1+r_{1}\right)$.
Setting $E=\alpha\mathtt{E}$ allows for the very simple input-output
relation $O=E-Y$. Using that for a resonant DBR $\phi_{1}\left(\bar{\omega}_{m}\right)=0$,
the coupling with the intra-cavity field $h=\alpha\tilde{h}$ simplifies
as 
\begin{align}
h= & \frac{(1-\left|r_{1}\right|)(1+\left|r_{2}\right|)}{1-\left|r_{1}r_{2}\right|}
\end{align}
We further define the cavity enhanced Kerr nonlinearity (in $\mathrm{m}^{2}/V^2$)
\begin{eqnarray}
s & = & \frac{3\bar{\omega}_{m}\bar{\chi}_{3}W}{8\bar{n}_{m}c}\frac{\left(1+\left|r_{2}\right|\right)\left(1+\left|r_{1}\right|\right)^{3}}{\left(1-\left|r_{1}r_{2}\right|\right)\left|t_{1}\right|^{2}},
\end{eqnarray}

Normalizing transverse space to the cavity-enhanced diffraction length
$l_{\perp}=\sqrt{\frac{L}{\bar{q}_{m}}\frac{\left|r_{1}r_{2}\right|}{1-\left|r_{1}r_{2}\right|}}$
and time by the photon lifetime $\kappa^{-1}$, and introducing the
effective detuning taking into account the slow thermal effects through
the temperature evolution
\begin{eqnarray}
\delta & = & \delta_{0}-2\bar{q}_{m}L\frac{\left|r_{1}r_{2}\right|}{1-\left|r_{1}r_{2}\right|}\frac{\partial_{T}\bar{n}_{m}}{\bar{n}_{m}}\delta T\label{eq:delta_tot}
\end{eqnarray}
leads us to the equation given in the main text 
\begin{eqnarray}
\dot{E} & = & \left[i\left(\Delta_{\perp}+s|E|^{2}-\delta\right)-1\right]E+hY,\label{eq:KGTI_clean}
\end{eqnarray}
We note that $s=+1$ (resp. $s=-1$) corresponds to a focusing (resp.
defocusing) Kerr medium while a complex value would take into account
two-photon absorption.

\paragraph*{Temperature equation}

We assume that the temperature of the micro-cavity relaxes exponentially
towards the equilibrium temperature $T_{0}$ set by the Peltier element
and at a rate $\gamma$. This thermalization is impeded
by the presence of a heat source. These ideas materialize as 
\begin{eqnarray}
\dot{T} & = & \gamma\left(T_{0}-T+R_{th}Q\right),
\end{eqnarray}
with $R_{\mathrm{th}}$ the thermal impedance in K/W and $Q$ in Watts.
The optical power injected into the system that is not re-emitted
from both sides of the micro-cavity must be dissipated as heat. This
simple statement allows finding that $Q$ is the differences between
the norm of Poynting vectors of the injected, reflected and transmitted
fields integrated over the surface of the device that we assume to
be circular with $R$, i.e.
\begin{eqnarray}
Q & = & \pi R^{2}\left\Vert \left\langle \boldsymbol{\mathcal{E}}_{i}\times\boldsymbol{\mathcal{H}}_{i}-\boldsymbol{\mathcal{E}}_{r}\times\boldsymbol{\mathcal{H}}_{r}-\boldsymbol{\mathcal{E}}_{t}\times\boldsymbol{\mathcal{H}}_{t}\right\rangle \right\Vert 
\end{eqnarray}
where the bracket $\left\langle \cdot\right\rangle $ represents an
average over a few oscillations of the electric field. Assuming no
field escapes by the bottom mirror (Gires-Tournois regime $\left|t_{2}\right|\rightarrow0$),
and plane waves so that $\boldsymbol{\mathcal{H}}=\boldsymbol{\mathcal{E}}/\eta$
with $\eta=\sqrt{\mu_{0}/\varepsilon_{0}}\sim377~\Omega$ the vacuum
impedance we find after simplifications that 
\begin{align}
\dot{\delta}= & \gamma\left[\delta_{0}-\delta-\mu_{th}\left(|Y|^{2}-|O|^{2}\right)\right]\label{eq:delta_therm}
\end{align}
with the coupling constant in $\mathrm{m}^{2}/V^2$
\begin{eqnarray}
\mu_{th} & = & R_{th}\frac{\pi R^{2}}{2\eta}2\bar{q}_{m}L\frac{\left|r_{1}r_{2}\right|}{1-\left|r_{1}r_{2}\right|}\frac{\partial_{T}\bar{n}_{m}}{\bar{n}_{m}}
\end{eqnarray}

Interestingly, we notice that our model does not include so far optical
absorption in the micro-cavity. Indeed, the term $-E$ in Eq.~\ref{eq:KGTI_clean}
is not \emph{per se} a loss term since it simply represents the coupling
to the external cavity. We can remedy to this problem modifying Eq.~\ref{eq:KGTI_clean}
with an additional dimensionless loss term $\alpha$ 
\begin{eqnarray}
\dot{E} & = & -\left(1+\alpha\right)E+i\left(s|E|^{2}-\delta\right)E+hY,\label{eq:KGTI_unclean}
\end{eqnarray}

Since thermal effects are slow, we can assume steady states operation
in Eq.~\ref{eq:KGTI_unclean}. In that case defining the nonlinear
detuning $\theta=\delta-s|E|^{2}$ we get $E=hY/\left(1+\alpha+i\theta\right)$
and the output $O=E-Y$ in the GTI limit where $h=2$ reads
\begin{eqnarray}
O & = & \frac{1-\alpha-i\theta}{1+\alpha+i\theta}Y
\end{eqnarray}
which corresponds to the intuitive result that the source term in
Eq.~\ref{eq:delta_therm} is proportional to the intracavity field
$\left|Y\right|^{2}-\left|O\right|^{2}=\alpha\left|E\right|^{2}.$
In the limit of small values of $\alpha$ and large $\mu_{th}$, one
can neglect $\alpha$ in Eq.~\ref{eq:KGTI_unclean}. We define $\mu_{\alpha}=\mu_{th}\alpha$
thereby reaching the model given in the main text. Interestingly,
we note that introducing nonlinear losses such as two-photon absorption,
i.e. $s\in\mathbb{C}$ would induce a dissipation term proportional
to the fourth power of the field amplitude.

The last step in our derivation consists in linking the injected field
$Y$ to the cavity output $O$, that is re-injected after one roundtrip
upon reflection on the external mirror with amplitude $\eta$ and
phase $\phi$, see Fig.~1. This proceeds without difficulty and taking
into account the continuous wave injection beam $Y_{0}$ we find 
\begin{eqnarray}
Y\left(t\right) & = & \eta e^{i\varphi}O\left(t-\tau\right)+\sqrt{1-\eta^{2}}Y_{0}
\end{eqnarray}
where we introduced the round-trip phase $\varphi=\omega_{0}\tau+\phi$.
In absence of transverse effects the final model reads
\begin{align}
\dot{E}= & \left[i\left(s|E|^{2}-\delta\right)-1\right]E+hY,\\
\dot{\delta}= & \gamma\left(\delta_{0}-\delta-\mu_{\alpha}\left|E\right|^{2}\right),\\
Y= & \eta e^{i\varphi}\left[E\left(t-\tau\right)-Y\left(t-\tau\right)\right]+\sqrt{1-\eta^{2}}Y_{0}.
\end{align}
\begin{figure}[b!]
	\centering
	\includegraphics[width=1\columnwidth]{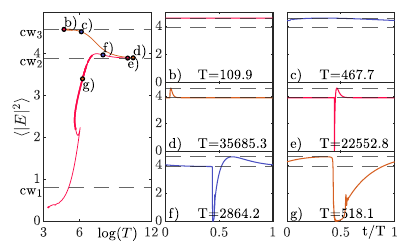}
	\caption{a) The left Canard region from Fig. \ref{fig:8} on the main text plotted over the period $T$. On the $\langle |E|^2\rangle$-axis one can see three continuous wave plateaus CW$_{1,2,3}$. The branch shows an incomplete homoclinic bifurcation, approaching CW$_2$. (b-g) Respective profiles from different marked parts of the branch.}\label{Supfig:1}
\end{figure}
\section{Extended bifurcation analysis}
\label{appendixB}
\begin{figure}[t!]
	\centering
	\includegraphics[width=1\columnwidth]{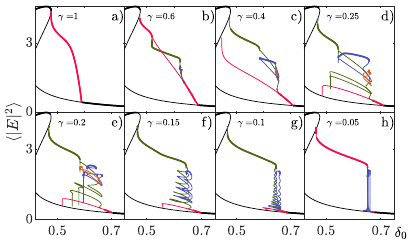}
	\caption{Emergence of mixed-mode oscillations and a sharp subthreshold to relaxation oscillation transition if a scale separation is introduced. In panels (a) to (h), $\gamma$ is gradually decrease, leading to a shattering of the periodic branch from (a) (red) towards the bifurcation scenario from Fig.~\ref{fig:8}, depicted in panel (h). Parameters as in Fig.~\ref{fig:8}.}\label{Supfig:2}
\end{figure}

%
Here, we provide additional insight into the complicated bifurcation scenario depicted in Fig.~\ref{fig:8}~(a). Starting with the bifurcation from the upper continuous wave (CW) value around iv), we observe a supercritical AH bifurcation, followed by a sharp canard transition. Using numerical path continuation on these solution, yields an interesting phenomenon in the period, seemingly diverging around the middle CW value. Notably, the branch approaches the middle CW value, without reconnecting to it, leading to an incomplete homoclinic bifurcation scenario as indicated in Fig.~\ref{Supfig:1}. Here, the relaxation oscillation branch from Fig.~\ref{fig:8}~(a) is shown with period $T$ as parameter on a logarithmic scale. Different profiles depiced in panels (b-g) indicate how the branch leaves the upper CW in a supercritical AH bifurcation ((b,c)), slowly approaching the middle CW value (d). After a cusp, a spike to zero appears (e), followed by the evolution of relaxation oscillations.
In general, the emergence of canard scenarios and mixed-mode oscillations (MMOs) was reported to be a phenomenon resulting from scale separation.
Hence, we show an exemplary bifurcation scenario, unveiling the path into scale separation. In Fig.~ \ref{Supfig:2}~(a), a periodic branch (red) is shown, emerging supercritically from a CW branch (black). The parameters are the same as in Fig.~\ref{fig:8}, except for the choice of $\gamma=1$. Reducing $\gamma$ leads to the emergence of a second periodic branch in a period doubling bifurcation, as well as additional small branches, which are the MMOs. As we further reduce $\gamma$, the scenario becomes more complex, while also getting confined to a decreasing parameter regime. Finally, in Fig. \ref{Supfig:2}~(h), we recover the scenario presented in Fig. \ref{fig:8}.
%
%

\end{document}